\documentclass[superscriptaddress,a4,pre,preprint]{revtex4}
\usepackage[OT1]{fontenc}
\usepackage{graphicx}
\usepackage{subfigure}
\usepackage{amsmath,amsfonts}
\usepackage{amstext,amssymb,amsbsy,amsthm,eucal}
\usepackage{latexsym}
\usepackage{wasysym}

\fontfamily{cmr}\selectfont 

\begin{document}

\author{Aonghus Lawlor, Paolo De Gregorio, Phil Bradley}
\affiliation{University College Dublin, Irish Centre for Colloid Science and Biomaterials, Department of Chemistry, Belfield, Dublin 4,
Ireland}

\author{Mauro Sellitto}
\affiliation{The Abdus Salam International Centre for Theoretical Physics, Strada Costiera 11, 34100 Trieste, Italy }

\author{Kenneth A. Dawson}
\affiliation{University College Dublin, Irish Centre for Colloid Science and Biomaterials, Department of Chemistry, Belfield, Dublin 4,
Ireland}

\title{Geometry of Empty Space is the Key to Near-Arrest Dynamics}

\begin{abstract}
We study several examples of kinetically constrained lattice models
using dynamically accessible volume as an order parameter. Thereby we
identify two distinct regimes exhibiting dynamical slowing, with a
sharp threshold between them. These regimes are identified both by a
new response function in dynamically available volume, as well as
directly in the dynamics. Results for the self-diffusion constant in
terms of the connected hole density are presented, and some evidence
is given for scaling in the limit of dynamical arrest.
\end{abstract}
\date{\today}
\email{aonghus@fiachra.ucd.ie} 
\maketitle

In nature, molecules, particles or other elements of a nearly arrested
system may sometimes stop moving without any accompanying sharp change
in the thermodynamic quantities. Such systems have not crystallized,
the free energy has not been minimized, and there is no obvious order
parameter for the `transition'. This phenomenon of precipitous, and
apparently collective, loss of motion has been named `dynamical
arrest' \cite{mezard2000,lawlor2002prl,berthier2003,jung2004,degregorio2004}, or jamming
\cite{liu1998,truskett2000,mehta2000}. It is involved in the `glass transition'
\cite{gotze1991} and even, apparently, gellation
\cite{dawson2001,dawson2002}.
In the field of glassification especially, very significant advances
have been made in developing our understanding
\cite{gotze1992,stillinger1995}. 
In the scientific community there is an emerging opinion that the many
modes by which complex condensed states are formed are all aspects of
the process of dynamical arrest.

We have sought a fundamental theoretical approach to dynamical arrest
\cite{lawlor2002prl,degregorio2004} based on a physical order
parameter to describe the vicinity of dynamical arrest, keeping in
mind the potential to make direct contact with experiment. In
particular, the advent of new experiments \cite{kegel2000,weeks2000}, 
and new experimental methods
\cite{scheffold2001,cheng2002,cipelletti2003} 
being developed in colloidal and soft matter science, opens up the
possibility to directly connect observation of particle configurations
and their fluctuations to suitably framed theoretical concepts. We
have identified dynamically available volume (DAV)
\cite{lawlor2002prl}, the ensemble of physical space available to
particle dynamics, as a possible order parameter. Its use has been
illustrated in a recent study of certain lattice models that exhibit
dynamical arrest
\cite{lawlor2002prl,degregorio2004,degregorio2005} and has
been considered in models of granular matter \cite{mehta2000}.

In this paper, we highlight a new insight that for the simple model
systems we explore, and possibly more generally in nature, the
geometry of available empty space of a nearly arrested system has a
profound impact on the nature of the relaxation processes and the
accompanying laws governing the onset of arrest. We propose that the
onset of arrest involves two qualitatively different regimes, both of
which appear to exhibit a degree of universality and scaling not
previously appreciated. Our observations are valid for a variety of
lattice types, and in both two and three spatial dimensions.

The models we study include kinetically constrained lattice models
introduced by Kob and Andersen \cite{kob1993} in which particles may
move to an adjacent empty site (a hole) if surrounded by $c$ or fewer
neighbors, and if the movement is reversible according to these same
rules. Vacancies are empty sites into which no adjacent particle can
move. A dynamical arrest transition occurs as a function of increasing
particle density, system size, spatial dimensionality $d$, and
constraints $c$.

At moderate densities a particle that moves into a hole typically
generates a new hole that may be used for a subsequent motion. By such
sequential motions, all particles in the system may be moved beginning
with such a (`connected') hole. At high density many (`disconnected')
holes are caged by a continuous boundary of particles that cannot be
broken by particle rearrangements inside the boundary (eg. in the
square-lattice KA model the boundary has at least a double row of
particles of rectangular shape). Each particle on this boundary is
blocked because it cages, and is caged by, its neighbors. In our
previous work we deemed disconnected holes irrelevant for the long
length and time scale transport coefficients, and the connected holes
(of density $\nu_c$) become the natural order parameter
\cite{lawlor2002prl}. Subsequently \cite{degregorio2004} it has become
apparent that, asymptotically close to final arrest (in the lattice
models, for $\rho=1$) this is equivalent, via the relation
$\nu_c=1/\xi^d$, to relating the transport coefficients to the
bootstrap length $\xi$ \cite{jackle1994,toninelli2004}.
\begin{figure}[t]
\begin{center}
\includegraphics[angle=-90,width=\columnwidth,height=!]{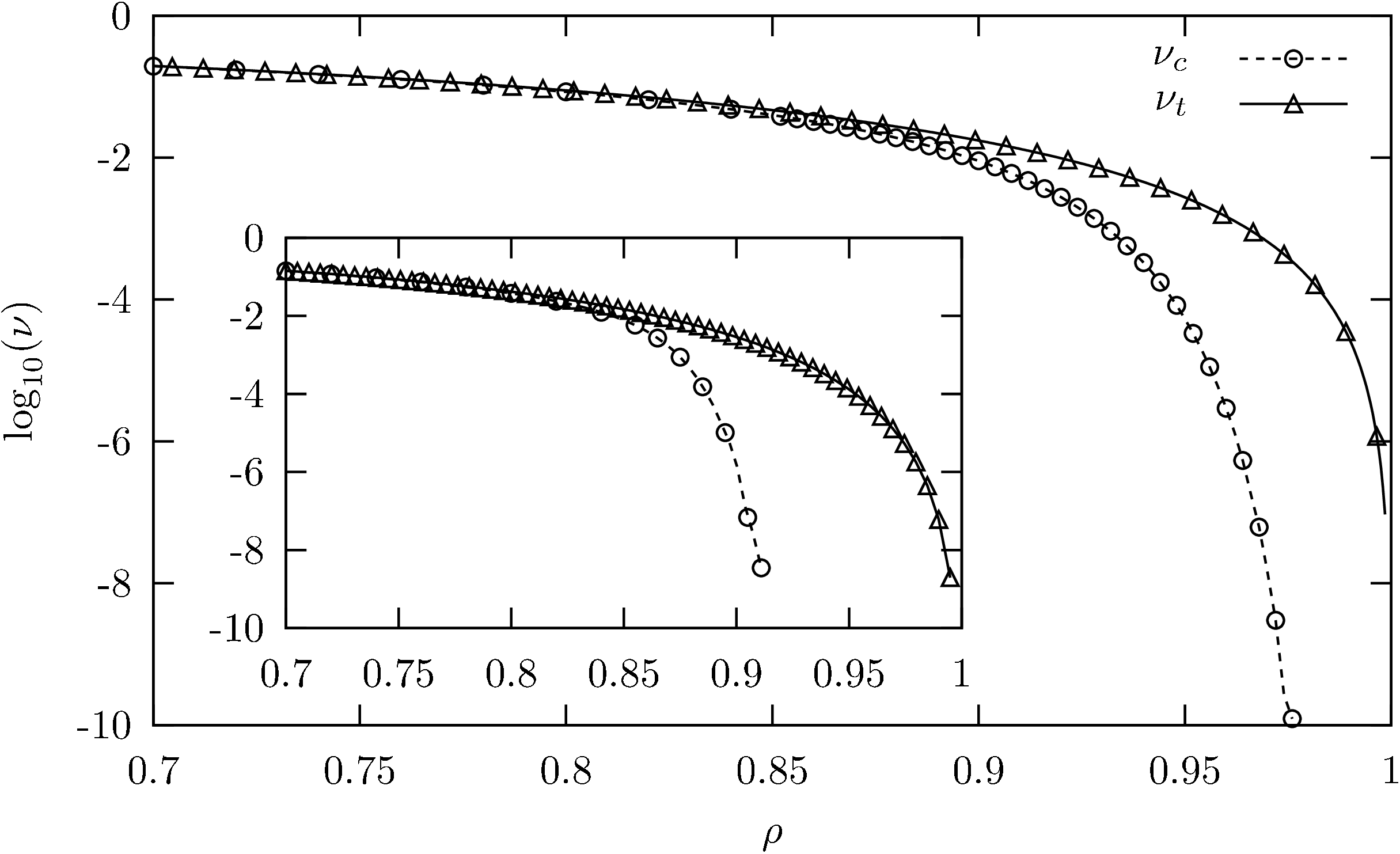}
\caption{
Total ($\nu_t$) and connected ($\nu_c$) hole density for the 2D and 3D
(inset) Kob-Andersen model- largest length corresponding to $\xi=10^5$
for 2D and $\xi=10^3$ for 3D.
} \label{fig:hole_density}
\end{center}
\end{figure}

Connected and disconnected hole densities can be estimated by
numerical simulation, or by theoretical approaches
\cite{degregorio2004}- in some two dimensional models they can be calculated exactly
\cite{degregorio2005}. The total hole density $\nu_{t}$ is a
sum of the connected $\nu_{c}$ and disconnected $\nu_{d}$ hole
densities, and $\nu_c$ is plotted for $c=d=2$ KA in
Fig. \ref{fig:hole_density} ($c=d=3$ KA in the inset). Although the
results shown here for the connected hole densities, or equivalently,
the bootstrap length $\xi^d = 1/\nu_c$ correspond to bootstrap
simulations for system sizes beyond any previously computed, it is
important to note that they remain far too small
\cite{degregorio2004,degregorio2005} to be described by simple
(exact) asymptotic approximations \cite{holroyd2003}.

\begin{figure}[t]
\begin{center}
\includegraphics[angle=-90,width=\columnwidth,height=!]{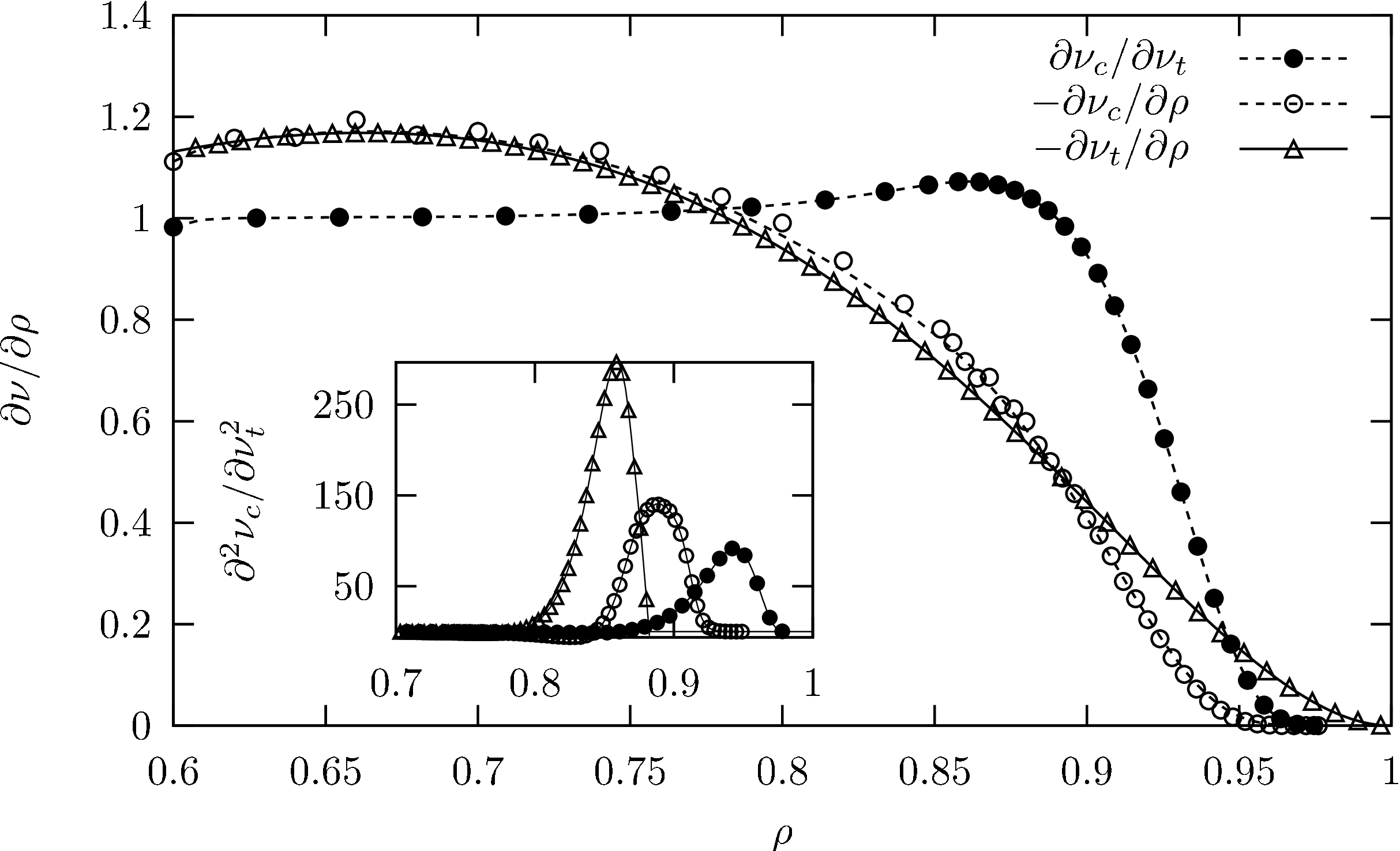}

\caption{ 
$\chi_c=\partial \nu_c/\partial \nu_{t}$ for $d=2$ KA model. Also
shown are $-\partial \nu_{c}/\partial \rho$ ($\circ$) and $-\partial
\nu_{t}/\partial \rho$ ($\triangle$). The inset shows $\eta_c
=\partial^{2} \nu_c/\partial \nu_{t}^{2}$ for cubic KA $c=3$
($\circ$), fcc KA $c=6$ ($\triangle$) and KA $c=d=2$ ($\bullet$).}

\label{fig_dnuc_dnut}
\end{center}
\end{figure}

The connected hole density is the order parameter of the dynamical
system, and determines the transport coefficients. The response of the
dynamical system to creation of a hole may be calculated from the
derivative (essentially a response function for hole creation)
$\chi_c=\partial \nu_c/\partial \nu_{t}$, plotted as a function of
density for the $c=d=2$ KA model in Fig. \ref{fig_dnuc_dnut}. The
result is striking- we find a threshold (`transition') density below
which new holes automatically become connected (and therefore
contribute to diffusion) and above which nearly all holes are trapped
within extended cages.

To the low density (`unstable') side of the transition a new hole is
typically close to an existing connected hole, and therefore likely
also to be connected. More holes lead to further mobility, and the
system is termed unstable. However, this mechanism also implies that
dynamical connectedness, formerly defined only in relation to phase
space (the possibility to move all particles from a given initial
condition) is strongly associated with geometrical aspects of a
network of connected holes in this regime.

To the high density (`metastable') side of the transition, creation of
a new hole typically leads to a rattler, or disconnected hole due to
caging. Existing connected holes are now dispersed sufficiently far
apart that they provide no assistance to a new hole.

The small peak in $\chi_c$ (at $\nu_c^*$) is considered the onset of
this transformation of phase space, and the peak of the second
derivative, $\eta_c=\partial^{2} \nu_c/\partial \nu_{t}^{2}$, an
effective threshold at which the system may be deemed to have made a
`transition' in dynamics (see inset to Fig. \ref{fig_dnuc_dnut}). In
that inset we show that a similar reorganization of phase space is
observable in models with different sets of kinetic constraints,
symmetry of lattice, and spatial dimensionality. It should be
emphasized that it has so far been difficult to quantitatively relate
this restructuring of phase space, and the change in dynamical
processes, to simple static concepts of connected-ness of vacancy and
hole networks, though attempts continue \cite{lawlor2005a}. On the
contrary, for each of the different models described in this inset
there are quite different spatially connected networks, whilst the
dynamical phenomenon is quite general. Note carefully that the
connected hole density contains information on the potential temporal
evolution of the system. We now consider the important step in
understanding the geometrical aspects involves the concept of response
functions. These are defined one level removed from the specific rules
of the model itself, that capture the spatio-temporal connected-ness
of phase space. This approach is expected to be more appropriate also
in attempts to generalize to the continuum.

We conclude there are likely two quite distinct dynamically slowed
regimes on approach to arrest, possessing qualitatively different
spatio-temporal or dynamical relaxation processes, different laws for
the diffusion constant, and non-linear response functions. There is
ample anecdotal experimental information for differing dynamical
regimes, and possibly differing types of law, but hitherto there has
been no clear formulation of the underlying ideas. We are in a
position to check this explicitly in the dynamics of these
models. Note however, that the strength of the calculations above are
that they are (or may be made) essentially exact, whereas the dynamics
calculations are lengthy with more limited reliability. We shall now
establish the existence of two different regimes for these models by
exploring different aspects of the dynamics.

\begin{figure}[]
\begin{center}
\includegraphics[angle=-90,width=\columnwidth,height=!]{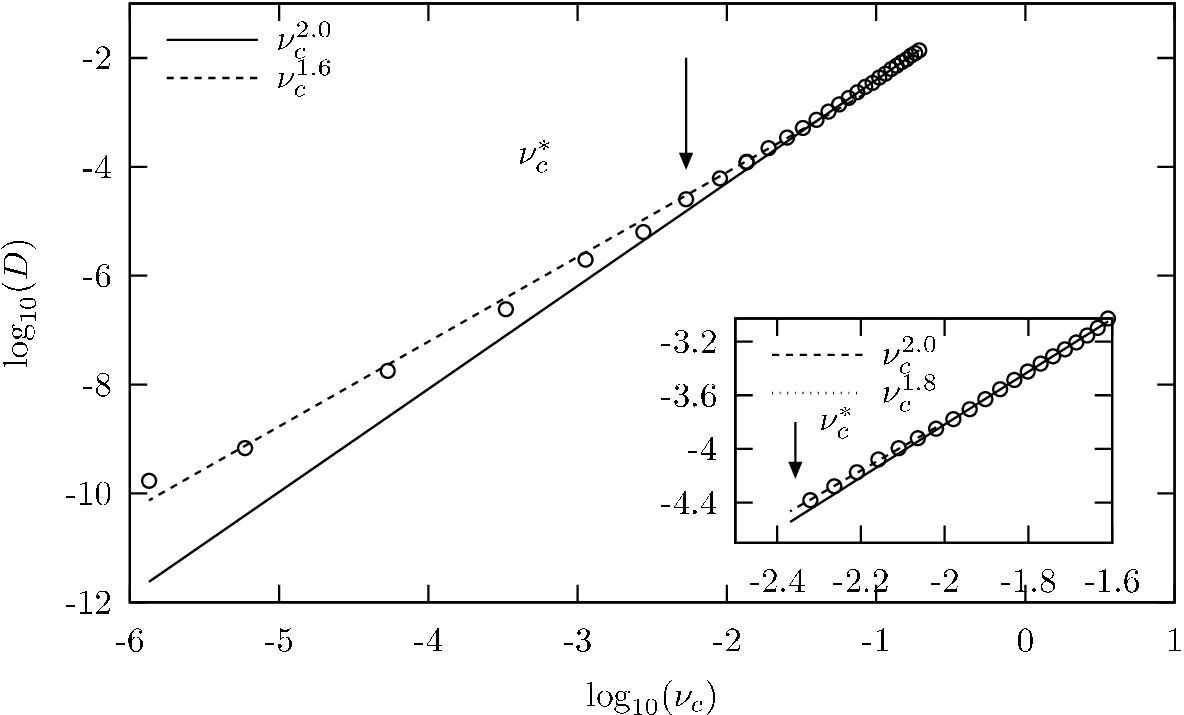}
\caption{Diffusion constants for 2D and 3D (inset) for KA} \label{fig:diffusion_constant}
\end{center}
\end{figure}

The mean squared distances travelled by particles $\left< r(t)^2
\right>$ are computed as usual \cite{lawlor2002prl}, and the diffusion
constants $D$ calculated from the long-time diffusivity $\left< r(t)^2
\right> = 2dDt$ (see Fig. \ref{fig:diffusion_constant}). Having
computed the relation between the connected hole density and particle
density we may eliminate the latter and plot the diffusion constant in
terms of the order parameter, $\nu_c$. It is known that the connected
hole density fails to reach a simple asymptotic form in any density
range accessible to computers
\cite{adler1990,adler1991,branco1999,kurtsiefer2003,degregorio2004,degregorio2005}. Indeed, in the
regimes we discuss, there is a difference of several orders of
magnitude from the asymptotic form that is usually assumed. Therefore,
in writing the diffusion constant in terms of the connected hole
density (or dynamical correlation or bootstrap length) it is important
to use results that are appropriate to the density regime under
consideration.

For moderate density we recover the known \cite{lawlor2002prl}
quadratic law $D = \gamma \nu_c^{2}$, ($\gamma$ is a model-dependent
effective rate constant) for different models, rules sets and spatial
dimensions. In that regime $\nu_c$ and $\rho$ (and therefore $D$ and
$\rho$) are connected by a power law.  However, at higher densities,
for those models that have been explored beyond the `transition',
there is a sharp change of slope and a new law for the diffusion
constant, $D = \gamma\nu_c^{z}$ ($z$ non-integer). In that regime the
hole density (and thereby diffusion constant) is connected to particle
density via an exponential-like decay, though the detailed form is
subtle \cite{degregorio2004}. In each case we have marked that value of
particle density at which the unstable-to-metastable transition takes
place based on available volumes. Examples are given for two and
three-dimensional KA models, but similar phenomena are present in
several other lattice types, and kinetic constraint sets. For the
three dimensional cubic lattice the cross-over is on the edge of the
presently accessible length and time scales.

One should not be complacent about quantitative treatment of this high
density limit as, for example, expressed in the (effective) exponents
$z$, which may contain a slowly varying density dependence. The fact
that the diffusion constants are derived from some of the longest
simulations and large system sizes is no reassurance; the dangers of
interpretation associated with the (related) subtle asymptotic
phenomena in the bootstrap percolation problem where system sizes far
beyond current computation fail to reach asymptotic laws is sufficient
warning
\cite{adler2003,holroyd2003,degregorio2004,degregorio2005}. Here we
will seek to clarify the distinct physical relaxation processes. The
true asymptotics may have to wait for developments in dynamics that
mirror those that have taken place recently in the bootstrap problem.
\begin{figure}[t]
\begin{center}
\begin{minipage}[t]{0.25\columnwidth}
{\includegraphics[width=0.8\columnwidth]{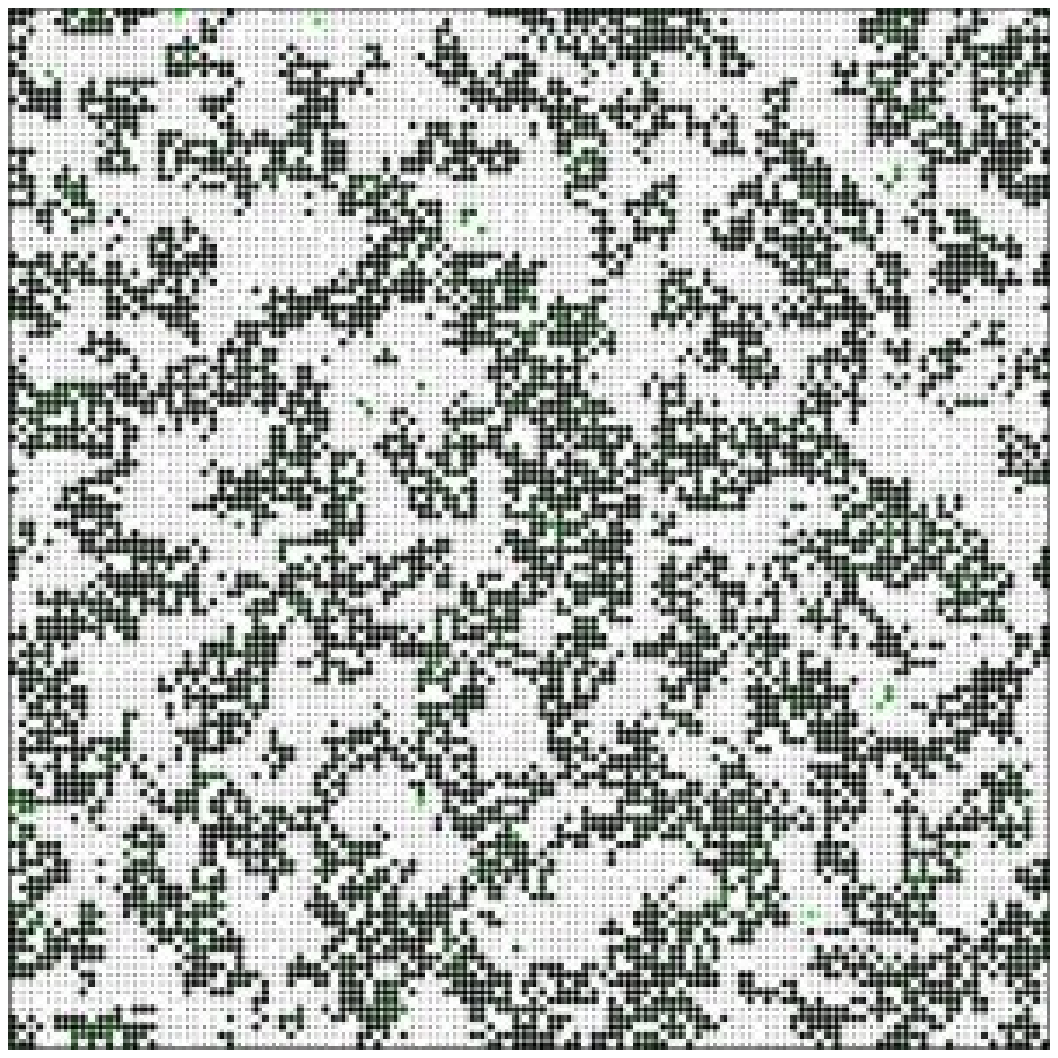}}
\end{minipage}
\hfill
\begin{minipage}[t]{0.25\columnwidth}
\includegraphics[width=0.8\columnwidth]{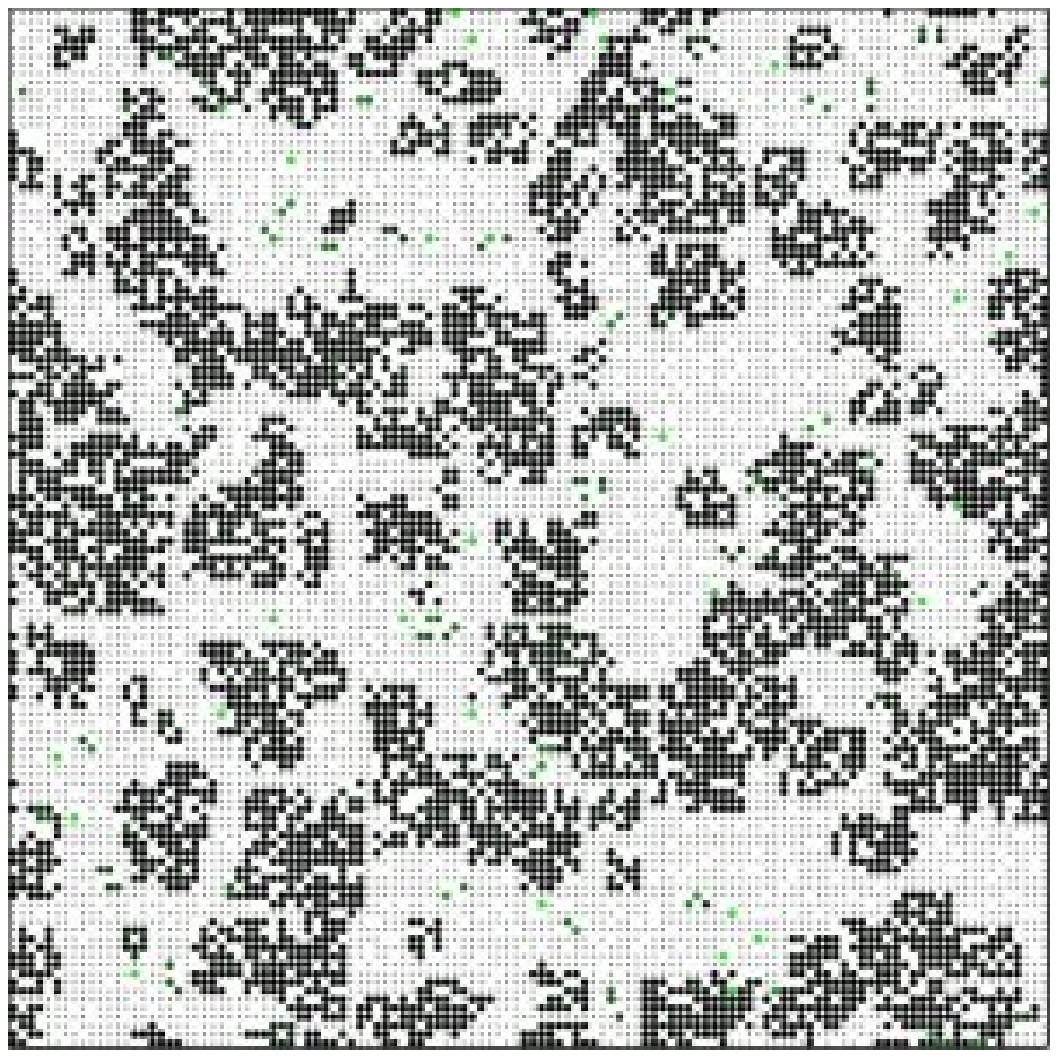}
\end{minipage}
\hfill
\begin{minipage}[t]{0.25\columnwidth}
\includegraphics[width=0.8\columnwidth]{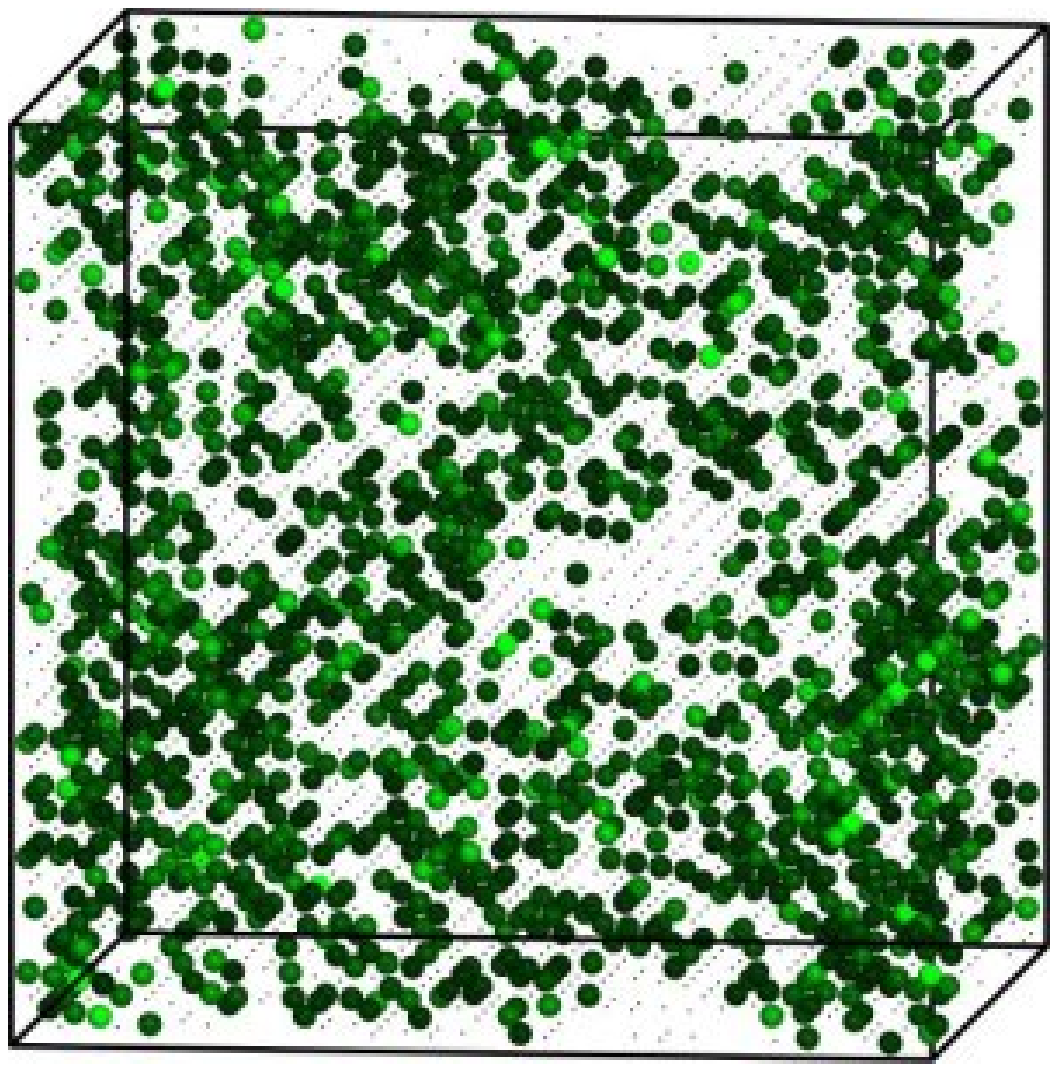}
\end{minipage}
\vspace{1pt}
\\
\begin{minipage}[t]{0.25\columnwidth}
{\includegraphics[width=0.8\columnwidth]{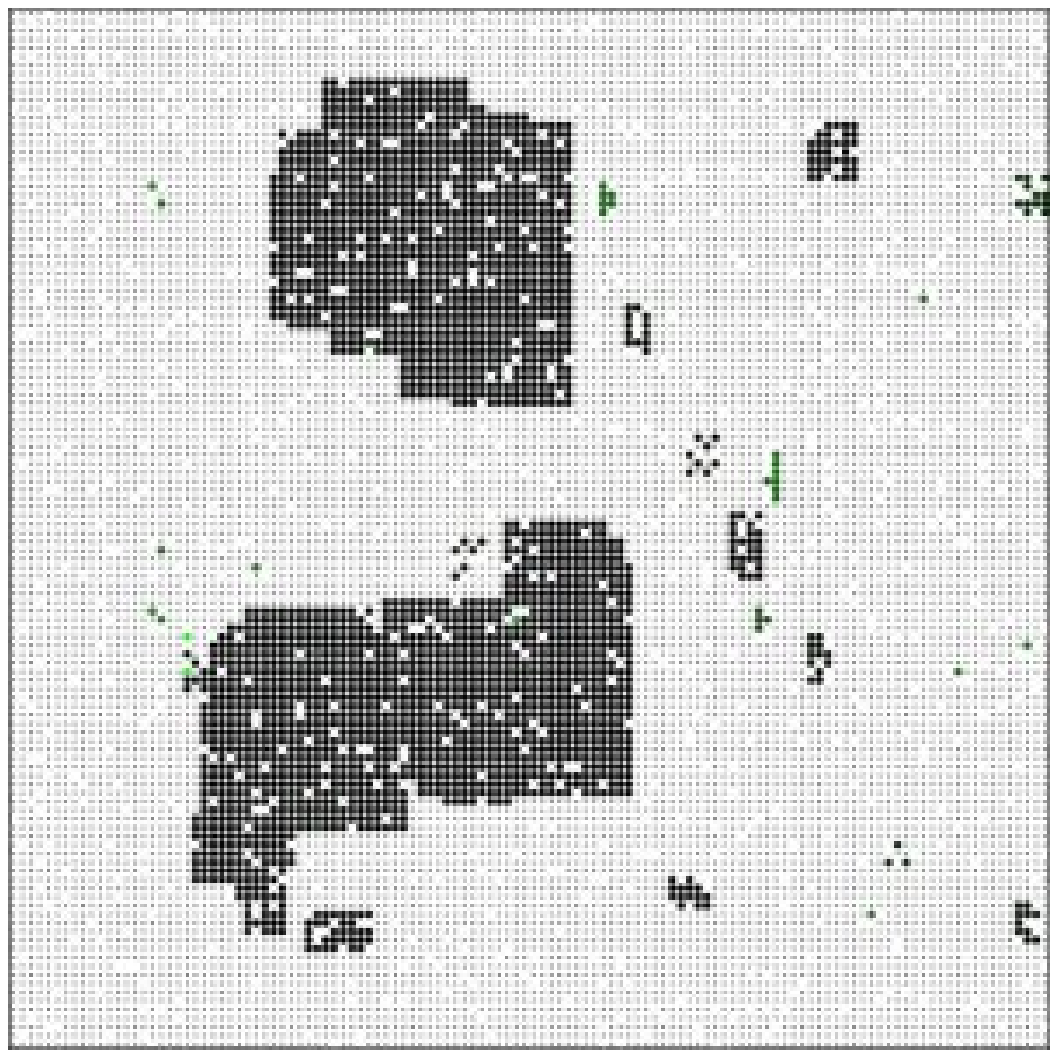}}
\end{minipage}
\hfill
\begin{minipage}[t]{0.25\columnwidth}
{\includegraphics[width=0.8\columnwidth]{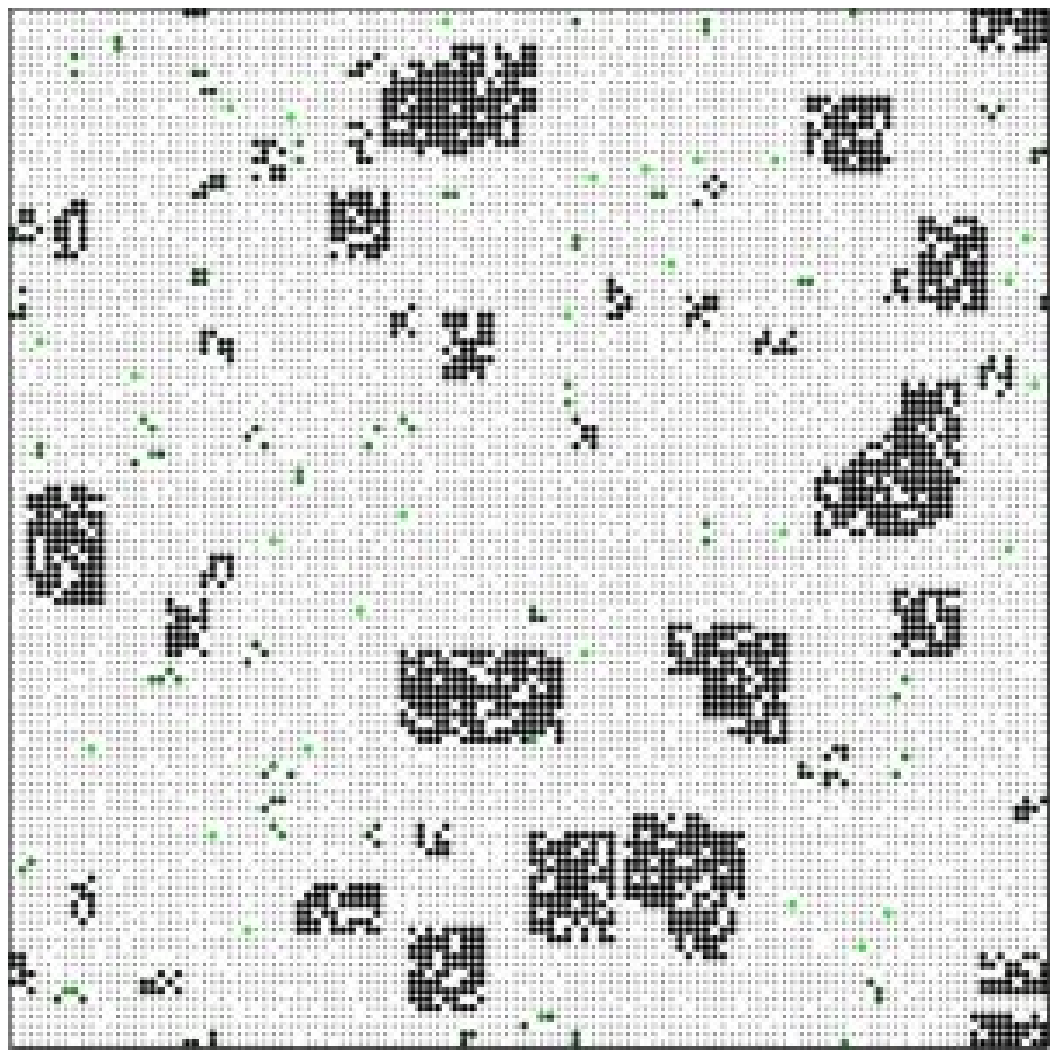}}
\end{minipage}
\hfill
\begin{minipage}[t]{0.25\columnwidth}
{\includegraphics[width=0.8\columnwidth]{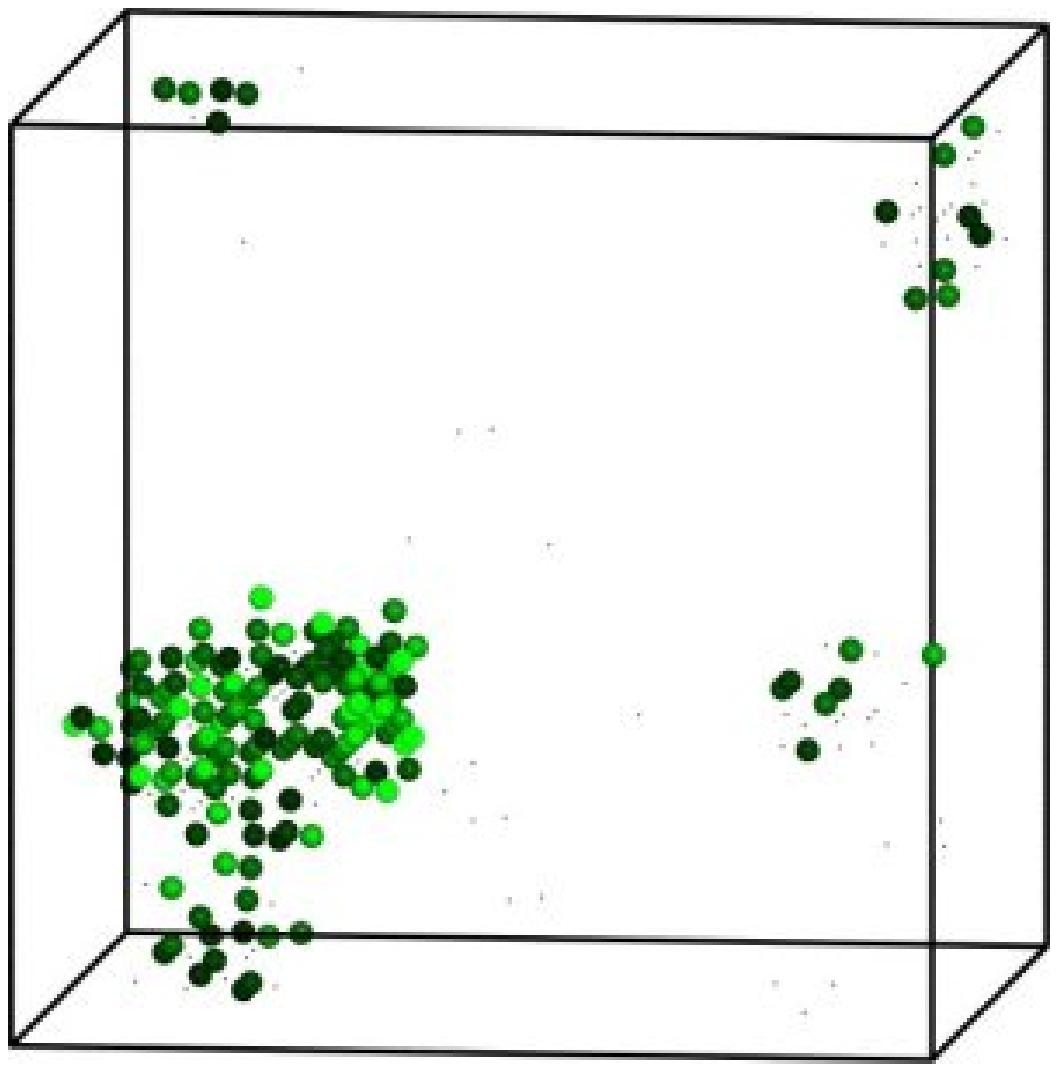}}
\end{minipage}
\\
\begin{minipage}[t]{0.29\columnwidth}
{\footnotesize{($a$) KA $c=2$, $\rho=0.82$, $0.94$}}
\end{minipage}
\hfill
\begin{minipage}[t]{0.29\columnwidth}
{\footnotesize{($b$) Modifed $c=2$, $\rho=0.80$, $0.87$}}
\end{minipage}
\hfill
\begin{minipage}[t]{0.29\columnwidth}
{\footnotesize{($c$) KA FCC $c=6$, $\rho=0.75$, $0.85$}}
\end{minipage}
\caption{
(Color online) Sample configurations of simulations of different models in the
unstable regimes (upper panels) and metastable regimes (lower
panels). We show only the particles that have moved after some
time. The patterns in the upper panels develop almost immediately,
while those on the bottom are essentially unchanged after many
millions of MCS. These give a pictorial representation of how
dynamically accessible volume is delocalized in the system, and
thereby characterize the nature of the relaxation processes in the two
regimes. In the unstable regime spinodal-like waves spread throughout
the system, whilst in the metastable system movement initiates in
localized droplets, and spreads slowly from those (rare) seeds.}
\label{fig:bootstrap_average}
\end{center}
\end{figure}

Amongst the most striking way to illustrate the consequences in
dynamics of the transition described in Fig. \ref{fig_dnuc_dnut}, is
to represent the spatio-temporal processes themselves visually. Thus
in Fig. \ref{fig:bootstrap_average} we illustrate qualitatively
different dynamics of unstable and metastable regimes from several
models using two representative densities on either side of the
geometrical transition.

The examples correspond to $c=d=2$KA and KA Modified and $c=6$, fcc KA
models (the $c=2$ KA Modified model is identical to the $c=2$ KA model
with the added restriction that when we consider a move, any of the
two vacant neighbours of the particle must be second neighbours to
each-other). In each case the particles which have moved after some
time are shown in their initial positions. We have explored many
examples, and found the phenomena we describe to be quite general. In
the unstable regime motion spreads rapidly in a concerted manner, with
ample pathways for configurational relaxation arising from implied
networks of connected holes. At higher particle densities, the
available volume disconnects (`de-percolates'), leading to the
metastable regime. There the pathways are quite different, the motion
is much slower and mediated by `droplets' within which connected holes
mobilize the particles with the assistance of a network of
vacancies. These pictures, illustrating the mechanisms by which
dynamically available volume is delocalized and configurational
relaxation occurs are quite consistent with the understanding of the
two regimes arising from the susceptibility ($\chi_c$) alone,
affirming the value and link between dynamics, and static averages in
available volumes, based on bootstrap processes.

\begin{figure}[t]
\begin{center}
\includegraphics[angle=-90,width=\columnwidth,height=!]{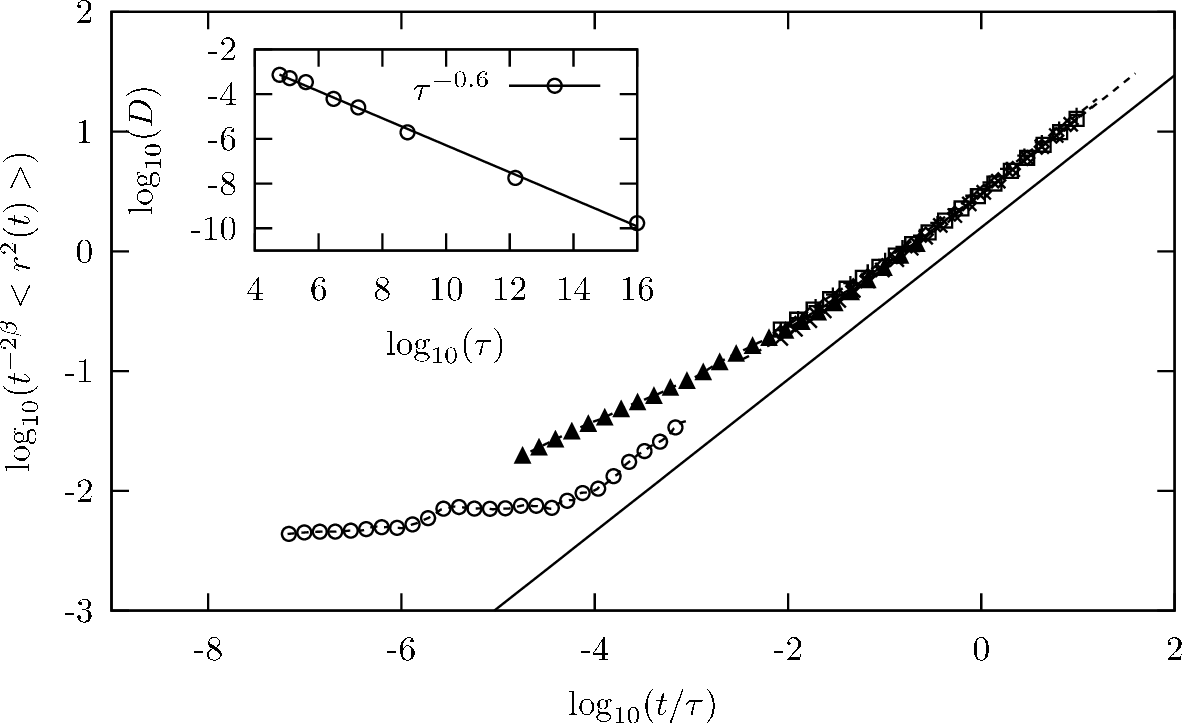}
\caption{
Scaling of the mean square displacement $\left< r^2 \right>=t^{2\beta} f(t/\tau)$
for 2D KA in the metastable regime. The densities are
$\rho=0.850$($\Diamond$), $\rho=0.860$($+$), $\rho=0.870$($\Box$),
$\rho=0.900$($\times$),
$\rho=0.930$($\blacktriangle$) at $L=512$ and $\rho=0.950$($\circ$) at
$L=1024$. In the inset we show the diffusion constant against
crossover time.
}
\label{fig:scaling}
\end{center}
\end{figure}
We end by noting that the metastable regime appears to exhibit a
characteristic time, and a degree of scaling, on approach to true
dynamical arrest. Thus, in Fig. \ref{fig:scaling} we show the scaled
curves, and in the inset the equivalent crossover time ($\tau$) at
which the system becomes diffusive. The scaling behaviour, and the
requirement for appropriate short and long-time behaviour leads us to
write $\left<r^2\right>=t^{2\beta} f(t/\tau)$ with $f(x) \sim const$
(small $x$) and $f(x) \sim x^\alpha$ (large $x$). Here $\beta$ is an
exponent reflecting the sub-ergodic nature of the system up to the
cross-over time \cite{rammal1983}. The two asymptotic behaviours match
at the crossover time, leading to $R^2=D\tau$, $R^2=\tau^{2\beta}$,
and therefore $D \sim \tau^{2\beta - 1}$. For large density the
connected hole density is written in terms of the bootstrap length,
and therefore we finally obtain the law, $D \sim \nu_{c}^{\mu}$ with
$\mu=-(2\beta-1)/\beta d$. From the cross-over data we estimate the
exponent $\mu=1.7$, in good agreement with direct measurements, and in
accord with recent studies of a similar model, the 2 vacancy assisted
triangular lattice gas \cite{pan2004}. The uncertainties of using
cross-over data are considerably larger than the direct measurement of
the exponent from the diffusion constant data above in
Fig. \ref{fig:diffusion_constant}, but the routes are entirely
independent, and therefore valuable in the context of the present
uncertainties about finite size dependence. We conclude that it is
possible that the cascade of cages and partial cages determining the
subergodic behaviour provides a set of scalable `traps', and thereby a
collapse of the dynamics data as outlined above.

In summary, we propose the existence of two regimes of near-arrested
dynamics with a transition between them driven by the underlying
geometry of available volume. The geometrical properties of this empty
space are not simple, but are characterized by a new response
function. The two regimes possess different relaxation processes,
leading to different dependencies of the diffusion constant on
density. That change in the diffusion constants could for some cases,
be so dramatic to be mistaken for the dynamical arrest itself. One
would expect to be able to observe the geometrical transition in
direct imaging experiments, and its consequences via new experimental
methods that can probe long-time and long length-scale dynamics in
colloidal and nano-particle science.

\begin{acknowledgements}
The work is supported by MCRTN-CT-2003-504712. Conversations with
G. Biroli, S. Franz and C. Toninelli are acknowledged.
\end{acknowledgements}


\end{document}